\documentclass[aps,prl,twocolumn]{revtex4}
\usepackage{graphicx}
\usepackage{dcolumn}
\usepackage{amsmath}
\usepackage{amssymb}

\def\ec{\epsilon_c}
\def\ex{\epsilon_x}
\def\exc{\epsilon_{xc}}
\def\be{\begin{equation}}
\def\ee{\end{equation}}
\def\bea{\begin{eqnarray}}
\def\eea{\end{eqnarray}}

\begin{document}
\title{Correlation energy and spin polarization in the 2D electron gas}
\author{Claudio Attaccalite, Saverio Moroni, Paola Gori-Giorgi, and
Giovanni B. Bachelet}
\affiliation{INFM Center for
  Statistical Mechanics and Complexity and
Dipartimento di Fisica, Universit\`a di Roma ``La Sapienza'', 
Piazzale A. Moro 2, 00185 Rome, Italy}
\date{\today}
\begin{abstract}
The ground state energy of the two-dimensional uniform electron gas
has been calculated with fixed-node diffusion Monte Carlo, including
backflow correlations, for a wide range of electron densities as a 
function of spin polarization.
We give a simple analytic representation of the correlation energy 
which fits the density and polarization dependence of the simulation 
data and includes several known high- and low-density limits.
This parametrization provides a reliable local spin density energy 
functional for two-dimensional systems and an estimate for the spin 
susceptibility. Within the proposed model for the correlation energy,
a weakly first-order polarization transition occurs shortly before 
Wigner crystallization as the density is lowered.
\end{abstract}
\maketitle

The two-dimensional electron gas (2DEG), realized in semiconductor
heterostructures~\cite{ando}, exhibits an extremely rich phenomenology
at low density, where correlations play an important role.  Being
outside the reach of perturbative approaches, many crucial aspects of
this interesting physics still lack a satisfactory
explanation~\cite{quali_aspetti}. Under these circumstances, valuable
information can be gained from simplified models, like the ideal 2DEG
(strictly 2D electrons interacting via a $1/r$ potential within an
uniform, rigid neutralizing background). At zero temperature, the
state of this system is entirely specified by the coupling parameter
$r_s=1/{\sqrt{\pi n}a_B}$, where $n$ is the density and $a_B$ the Bohr
radius, and the spin polarization $\zeta$.
Even for such a simple model the theory has greatly benefitted from 
numerical work, resulting --at least for a limited range of physical 
properties and/or system sizes-- in benchmark results, input quantities 
for approximate theories, and aspects of the ground--state phase diagram. 
The Quantum Monte Carlo (QMC) method~\cite{mitas}, for example, has accurately
predicted~\cite{TC,rapisarda} the critical density for Wigner crystallization 
of the ideal 2DEG, later observed experimentally~\cite{yoon}, and has
characterized~\cite{varsano} the polarization transition as being
weakly first-order and occurring shortly before crystallization as the
density is lowered.

In this work we present extensive QMC simulations of the liquid phase 
within the whole range of density and polarization, and provide an
analytic expression for the correlation energy $\epsilon_c$ as a
function of $r_s$ {\em and} $\zeta$.  Previous estimates of the
$\zeta$ dependence were based on interpolation conjectures between the
energy of the paramagnetic and of the fully polarized
liquid~\cite{TC,usanoTCePZlike}, but, as shown later, 
these estimates may significantly depart from the
calculated energy at intermediate polarization.

The interest in the $\zeta$ dependence stems from its use in the
spin-density functional theory (SDFT) of two-dimensional
systems~\cite{usanoTC,usanoTCePZlike,DFT2D}, in the development of
density functionals in the presence of magnetic fields~\cite{TCperB0},
and, more generally, in the study of the spin-polarization 
transition, whose role on transport properties~\cite{quali_aspetti} 
is prompting intense experimental investigation\cite{vitalkov}.

Our combination of numerical results and known analytic properties 
yields an estimate of the paramagnetic susceptibility. At low density, 
this is an extremely difficult quantity to evaluate with
approximate theories~\cite{tosi} due to the need of disentangling the
tiny --but important-- effect of quantum statistics from the huge
effect of correlation.
Finally, since we include the effect of backflow correlations\cite{kwon} 
as explained below, our $\epsilon_c(r_s,\zeta)$ provides a more 
accurate phase diagram than previously reported~\cite{TC,rapisarda}. 
Hartree atomic units are used throughout this work.

{\it Numerical results --}
Our calculations use standard fixed--node diffusion Monte Carlo
(FN-DMC)~\cite{mitas}, which projects the lowest-energy eigenstate
$\Phi$ of the many-body Hamiltonian with the boundary condition that
$\Phi$ vanishes at the nodes of a trial function $\Psi$. 
The algorithm simulates the imaginary-time evolution by a branching random
walk of $M$ copies of the $N$-electron system, using a short-time
approximation of the importance-sampled Green's function~\cite{mitas}. 
For each of the densities corresponding to $r_s=1, 2, 5, 10$ we have
considered about 20 values of $N$\cite{notaN} and 10--12 polarizations
$\zeta=(N^\uparrow-N^\downarrow)/N$. 
A few simulations have been done also for $r_s=40$, $\zeta=1$.
The electrons are placed in a square box with periodic boundary
conditions. Long-ranged interactions are dealt with by a model
potential (Eq.~(68) of Ref.~\cite{fraser})
which has been shown to give smaller finite-size corrections
than Ewald sums for the electron gas.
The bias introduced by the
finite population $M$ of walkers and the finite time step $\tau$ was
evaluated for selected systems and interpolated for all the others,
according to standard procedures. To estimate the difference $\Delta$
between the energy $\epsilon_N(r_s,\zeta)$ of the finite system and
its thermodynamic limit $\epsilon(r_s,\zeta)$, defined as
$N\to\infty$ at fixed density, we adopted a less usual strategy. 
Rather than a separate size extrapolation for each density
based on variational energies~\cite{TC,rapisarda,kwon}, we performed
a global fit directly based on FN-DMC energies, which exploits two
physically motivated ingredients: (i) the Fermi-liquid-like size
correction~\cite{ceperley78}
\bea \Delta(r_s,\zeta,N) = && \,\, \epsilon_N(r_s,\zeta) -
\epsilon(r_s,\zeta) = \nonumber\\
\delta(1 + \lambda\zeta^2) 
[ && t_N(r_s,\zeta)-t_s(r_s,\zeta)]-{(\eta+\gamma\zeta^2)}/{N},
\label{delta}
\eea
[$t_N$ is the energy per particle of $N$ noninteracting
electrons confined in a 2D box of side $L=(N/n)^{1/2}$ with
periodic boundary conditions, $t_s=(1+\zeta^2)/2r_s^2$ is
equal to $\lim_{N\to \infty} t_N$ with $n$ kept fixed, and
$\delta, \lambda, \eta, \gamma$ are
$r_s$-dependent parameters]; (ii) an
analytic expression for $\epsilon(r_s,\zeta)$, detailed below,
which involves 12 more free parameters.  On these grounds all the
FN-DMC energies $\epsilon_N(r_s,\zeta)$ calculated here for 
$r_s=1$ to $10$, plus those of Ref.~\cite{varsano} for $r_s=20$ 
and $30$, plus the energy (independently extrapolated to the 
thermodynamic limit) for $r_s=40$, $\zeta=1$ --a total of 122 
data-- formed the input for a best fit of the 36 free parameters, 
24 of which disappear from the final analytic expression since they 
only concern the size extrapolation. This fit yields a reduced 
$\chi^2$ of $3.8$.  More details will be reported elsewhere.

The fixed-node approximation is variational in character~\cite{mitas}, and
its accuracy depends on the nodal structure of $\Psi$. 
We choose a  Slater-Jastrow trial function
$\Psi(R) = J(R) D^\uparrow(R)D^\downarrow(R)$ where 
$R\!=\!\{{\bf r}_1,\ldots,{\bf r}_N\}$ represents the electronic
coordinates, $J(R)=\exp \left [ \sum_{i,j} u_{s_i,s_j} (r_{ij}) \right]$
is a symmetric Jastrow factor with different pseudopotentials $u_{s_i,s_j}$
for like-- and unlike--spin pairs~\cite{OB}, and $D^{\uparrow(\downarrow)}$ 
is a Slater determinant for spin-up(down) electrons.

The standard choice with homogeneous systems is to use plane waves
(PW) as single-particle orbitals:
$D_{PW}^{\uparrow(\downarrow)}=\det\left[\exp(i{\bf k}_i\cdot{\bf
r}_j)\right]$\cite{notaN}. However, within the fixed--node approximation, better
results are obtained with backflow (BF) correlations in the wave
function~\cite{kwon},
$D_{BF}^{\uparrow(\downarrow)}=\det\left[\exp(i{\bf k}_i\cdot{\bf
x}_j)\right]$, with ${\bf x}_i={\bf r}_i+\sum_{j\neq
i}^N\eta_{BF}(r_{ij})({\bf r}_i-{\bf r}_j)$. Since simulations with
the BF wave function are considerably more demanding than with PW
determinants, we calculated BF energies only for $\zeta\!=\!0,
N\!=\!58$ and $\zeta\!=\!1, N\!=\!57$ for each density, including
correlation functions $\eta_{BF}(r)$ as described in Ref.~\cite{kwon} 
optimized with variational Monte Carlo simulations\cite{mitas} in each case.  
The expected error with BF nodes is much smaller
than the difference between the PW and BF energies. For $r_s=1$, an
exact calculation\cite{kwon2} shows agreement with the BF result within
the statistical error.
The BF results are compared with PW energies in Table~\ref{tab_dmc}.
For other values of
$N$ and $\zeta$ the effect of backflow is estimated as a quadratic
interpolation in $\zeta$ and appended to PW energies, under the
further assumption that the size dependence be the same for BF and PW:
the $\sim 120$ FN-DMC data used in the fit of Eq.~(\ref{delta})
are the calculated PW energies, corrected for the time-step and population
bias, and shifted by the estimated backflow effect --the
infinite-size extrapolation being given as a result of the fit.

As expected~\cite{varsano}, BF correlations lower the energy more in the
paramagnetic than in the polarized phase. At large $r_s$ this relative
gain of the paramagnetic phase is a
significant fraction of the difference in energy between the two phases. 

\begin{table}[t]
{\centering \begin{tabular}{ccccc}
\hline\hline
 & \multicolumn{2}{c}{PLANE WAVES }& \multicolumn{2}{c}{BACKFLOW }\\
\hline 
\( r_s  \)& \( \zeta =0 \)& \( \zeta =1 \)& \( \zeta =0 \)& \( \zeta =1 \)\\
\hline 
1&-0.2013(1) &0.13147(2) & -0.20372(4) & 0.13109(4) \\
2&-0.255802(4) &-0.193349(1) &-0.25721(3) &-0.19359(2) \\
5&-0.149134(9) &-0.143520(5)  &-0.149518(9) & -0.143610(7)\\
10&-0.0852706(4) &-0.084555(2) & -0.085427(6) &  -0.084584(2) \\
20&-0.046241(1) &-0.0462385(6) &-0.046283(1) & -0.0462488(8)  \\
30&-0.031923(1) & -0.0319298(6) &-0.031941(2) & -0.031938(1) \\
\hline\hline 
\end{tabular}\par}
\caption{{FN-DMC energies with plane wave or backflow nodes.
Data pertain to simulations with $N=58$ for $\zeta=0$ and $N=57$ for $\zeta=1$,
$M=200$ and $\tau=0.002$, $0.01$, $0.1$, $0.3$, $1.0$, $2.0$ in order of 
increasing $r_s$.
}}
\label{tab_dmc}
\end{table}

\begin{table}[t]
\begin{tabular}{lccc}
\hline\hline
 &  $i=0$ & $i=1$ & $i=2$  \\
\hline
 $A_i$ &$-0.1925^*$    & $0.117331^*$              &$0.0234188^*$   \\  
 $B_i$ & $0.0863136^*$ &$-3.394 \times 10^{-2} $    &$-0.037093^*$  \\  
 $C_i$ & $0.0572384$    &$-7.66765\times 10^{-3*}$             &$0.0163618^*$\\ 
 $E_i$ & $1.0022 $      &  $0.4133 $                  &$1.424301$      \\  
 $F_i$ & $-0.02069$    & $0^*$                     & $0^*$       \\  
 $G_i$ & $0.33997$       & $6.68467\times10^{-2}$       & $0^*$    \\  
 $H_i$ & $1.747\times 10^{-2}$      & $7.799\times 10^{-4}$  & $1.163099$   	 \\

\hline\hline  
\end{tabular}
\caption{{Optimal fit parameters for the correlation
energy, as parametrized
in Eqs.~(\ref{eq_exczdep}) and~(\ref{eq_alpha}). Values
labelled with $^*$ are obtained from exact conditions.
The parameter $D_i=-A_iH_i$ is not listed (see text); the parameter
$\beta$ is equal to 1.3386.}}
\label{tab_ecinf}
\end{table}

{\it Analytic representation for $\epsilon_c(r_s,\zeta)$ --} 
We parametrize the energy $\epsilon(r_s,\zeta)$ as follows.  We first
noticed that the spin-polarization dependence of the
exchange-correlation energy $\exc=\epsilon-t_s$, as given by the DMC
data, is very well described by the simple form
$c_0+c_1\zeta^2+c_2\zeta^4$ for $r_s\gtrsim 5$.  On the other hand,
the known high--density limit~\cite{mike,HD}, 
\begin{equation}
\exc(r_s,\zeta) =  \ex(r_s,\zeta)+
a_0(\zeta)+b_0(\zeta)r_s\ln r_s+O(r_s),
\label{expansion}
\end{equation}
contains non-negligible contributions from higher powers of $\zeta$:
the dominating exchange term $\epsilon_x$ is equal to
$-2\sqrt{2}[(1+\zeta)^{3/2}+(1-\zeta)^{3/2}]/3\pi r_s$.  
Since we want to interpolate the
energy between high and low density, we choose a functional form 
for the correlation energy $\ec=\exc-\ex$ which
quenches the contributions from $\epsilon_x$ beyond fourth order in
$\zeta$ as $r_s$ increases,
\bea
\ec(r_s\zeta) & = & (e^{-\beta r_s}-1) \ex^{(6)}(r_s,\zeta)\nonumber \\
& + &\alpha_0(r_s)+\alpha_1(r_s)\zeta^2+\alpha_2(r_s)\zeta^4,
\label{eq_exczdep}
\eea
where
$\ex^{(6)}(r_s,\zeta)=\ex(r_s,\zeta)-
(1+\tfrac{3}{8}\zeta^2+\tfrac{3}{128}\zeta^4)\ex(r_s,0)$
is the Taylor expansion of $\epsilon_x$ beyond fourth order in
$\zeta$. Since the first term in the rhs of Eq.~(\ref{eq_exczdep})
contains powers $6$ and higher of $\zeta$, it is
immediate to identify the function $\alpha_0(r_s)$ as the correlation
energy at zero polarization,
$\alpha_0(r_s)=\epsilon_{c}(r_s,0)$. 
Furthermore, $2\,\alpha_1(r_s)=
\left[\partial^2\ec(r_s,\zeta)/\partial\zeta^2\right]_{\zeta=0}$
(spin stiffness), and $24\,\alpha_2(r_s)=
\left[\partial^4\ec(r_s,\zeta)/\partial\zeta^4\right]_{\zeta=0}$~. 
Our choice for the functions $\alpha_i(r_s)$ is a generalization of
the Perdew--Wang~\cite{PW92} form to the 2D case,
\bea
\alpha_i(r_s) & = & A_i+(B_ir_s+C_ir_s^2+D_ir_s^3)\times \nonumber \\
& & \ln\left(1+\frac{1}{E_ir_s
+F_ir_s^{3/2}+G_ir_s^2+H_ir_s^3}\right),
\label{eq_alpha}
\eea
which features both the subleading contributions of the
expansion (\ref{expansion}) and terms in $r_s^{-1}$ and $r_s^{-3/2}$
for $r_s\to\infty$\cite{bonsall}.
With suitable constraints, which leave only 12 independent parameters
in Eqs.~(\ref{eq_exczdep}) and (\ref{eq_alpha}), our correlation energy
satisfies exactly several known high-density and low--density limits:
(i) the exact values\cite{HD,mike} of $a_0(\zeta)$ and $b_0(\zeta)$ at
$\zeta=0$ and $\zeta =1$ in the small--$r_s$ expansion of $\epsilon_c$
shown in Eq.~(\ref{expansion}); (ii) the vanishing of the correlation
energy in the $r_s \to \infty$ limit, which simply implies that
$D_i=-A_iH_i$; (iii) the requirement that $\epsilon(r_s,\zeta)$ be
independent of $\zeta$ for $r_s \to \infty$.  The latter condition is
enforced up to order $O(r_s^{-2})$, thus recovering the low-density
behavior $\epsilon \to - m/r_s + n/r_s^{3/2} + O
(r_s^{-2})$\cite{bonsall} with positive $m$ and $n$ independent of
$\zeta$.
\begin{figure}[t]
\includegraphics[width=\columnwidth]{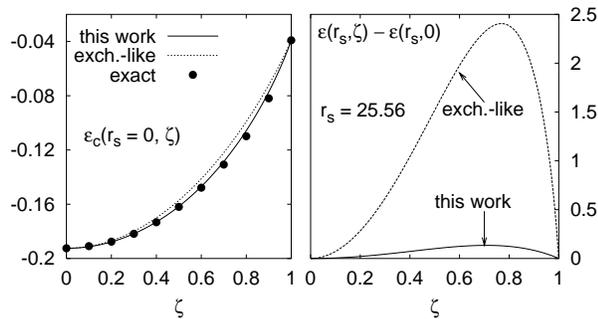} 
\caption{{Left panel: $\zeta$-dependence of the correlation energy in the
high-density limit.
The result of the present work is compared with the 
exact values from Ref.~\cite{mike} and with the exchange-like
interpolation of Refs.~\cite{usanoTCePZlike}. Right panel:
$\zeta$-dependence of the total energy at the polarization
transition density, $r_s=25.56$. The value
at $\zeta=0$ has been subtracted, and the result multiplied by
$10^5$.} Our result
is compared with the exchange-like interpolation. }
\label{fig_a0z}
\end{figure}

The optimal values of the parameters are listed in 
Table~\ref{tab_ecinf}.  In
the high-density limit, we can compare the $\zeta$-dependence of our
correlation energy with the exact one~\cite{mike}, finding very good
agreement, as shown in the left panel of Fig.~\ref{fig_a0z}, where the
widely-used exchange-like interpolation~\cite{usanoTCePZlike} is
also reported. In the right panel we instead see the
$\zeta$-dependence of the total energy at $r_s=25.56$
where, according to our results, the transition to the
fully polarized gas occurs. At such low
densities, the exchange-like interpolation scheme 
(here performed using our energy values at $\zeta=0$ and $\zeta=1$) 
significantly deviates from the QMC result. Both predict a
sudden transition from the 
$\zeta=0$ to the $\zeta=1$ fluid, but the energy barrier between
the two phases given by QMC is more then an order of magnitude smaller,
which reflects in a very large value of the spin susceptibility.

{\it Spin susceptibility --} In our parametrization, the spin
susceptibility $\chi$ of the ideal 2DEG is simply:
\begin{equation}
\frac{\chi}{\chi_0}=
\left[1-\frac{\sqrt{2}}{\pi}r_s+2\,r_s^2\,\alpha_1(r_s)
\right]^{-1},
\label{eq_ss}
\end{equation}
where $\alpha_1$ is given by Eq.~(\ref{eq_alpha}) and
Table~\ref{tab_ecinf}. As mentioned, it turns out that $\chi$ 
is a very delicate quantity, so that different theoretical 
predictions significantly depart from each other at $r_s$ as 
low as 3 or 4~\cite{tosi}.
  
\begin{figure}[t]
\includegraphics[width=7cm]{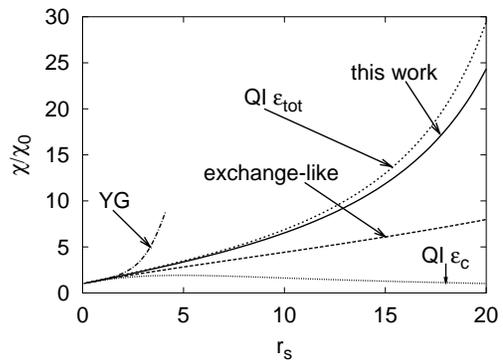} 
\caption{{
Spin-susceptibility for the 2DEG as a function of $r_s$.
The present result is compared with the exchange-like
approximation~\cite{usanoTCePZlike}, with two
quadratic interpolations~\cite{TC} based on the total
energy (QI$_{\epsilon_{tot}}$) and on the correlation energy 
(QI$_{\ec}$), and
with the Yarlagadda and Giuliani (YG)~\cite{tosi} calculation.}}
\label{spinsusc}
\end{figure}

In Fig.~\ref{spinsusc} we compare our result with other estimates
of $\chi$.
The result of the Yarlagadda and Giuliani (YG) calculation~\cite{tosi}
blows up already at $r_s\sim 4$. The tendency to predict a polarization 
transition at too high densities is shared by several approximate 
theories\cite{tosi}.
The exchange-like interpolation~\cite{usanoTCePZlike} is significantly
lower than the QMC estimate, as expected from Fig.~\ref{fig_a0z}.
Out of two recipes given by Tanatar and Ceperley (TC)~\cite{TC}, based 
on a quadratic interpolation of the $\zeta$-dependence of the total 
energy (QI$_{\epsilon_{tot}}$) and of
the correlation energy (QI$_{\ec}$) respectively, the former
(which yields by definition a divergent $\chi$ at the
polarization density) is quite accurate for $r_s\lesssim 10$,
whereas the latter greatly underestimates the
spin-susceptibility at all but the smallest $r_s$ (here the
quadratic interpolation has been performed using our energy values
at $\zeta=0$ and $\zeta=1$).

Our spin susceptibility is related to the
second derivative of the model correlation energy, which incorporates
an optimal interpolation of the QMC results and is constrained by
known limiting behaviors at very low and very high densities.
For this reason it represents a 
considerable progress over existing theories, providing a 
sound reference for further studies.

Nevertheless, the precise value of $\chi$ at very large $r_s$
(say, above 20) has to be taken with some caution.
When the  $\zeta$-dependence of $\epsilon(r_s,\zeta)$ 
is extremely weak (see the lower curve in the right 
panel of Fig.~\ref{fig_a0z}) the calculation of its derivatives
is beyond the accuracy of the present calculation, due to the 
fixed-node bias, the assumption of quadratic contributions from 
BF correlation, possible uncertainties from  the size extrapolation, 
statistical noise, and the chosen functional form of $\ec$, 
Eq.~(\ref{eq_exczdep}). The very nature of the polarization 
transition, wealky first-order according to Fig.~\ref{fig_a0z}, 
is clearly based on the above approximations and assumptions. 
\begin{figure}[t]
\includegraphics[width=7cm]{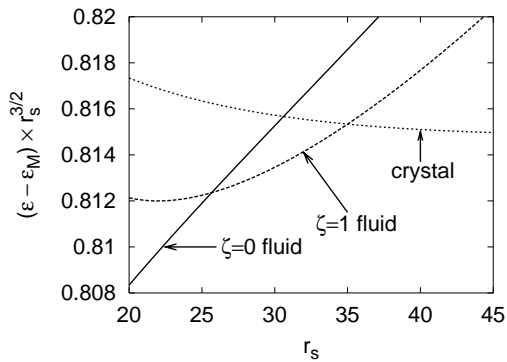} 
\caption{{$r_s$-dependence of the ground-state energy $\epsilon$ 
of the 2DEG for the
paramagnetic ($\zeta=0$) and ferromagnetic ($\zeta=1$) fluid phases, and 
for the Wigner crystal as given in Ref.~\cite{rapisarda}. 
$\epsilon_M=-1.1061/r_s$ is the Madelung energy.}}
\label{phasedia}
\end{figure}

{\it Phase diagram --} The above results allow us to draw the
zero--temperature phase diagram shown in Fig.~\ref{phasedia}, relative
to the Wigner crystal (WC), the paramagnetic liquid (PL) and the
ferromagnetic liquid (FL). Intermediate-polarization phases never
represent the stable phase in 2D (Ref.~\cite{varsano} and 
Fig.~\ref{fig_a0z}). For the fluid
phases, the energy $\epsilon=t_s+\epsilon_{xc}$ is given by
Eq.~(\ref{eq_exczdep}) with the parameters of Table~\ref{tab_ecinf}. 
The energy of the crystal is taken, instead, from Ref.~\cite{rapisarda}, 
since neither backflow nor spin polarization play a significant role in the
solid phase~\cite{nota}.

Similar FN-DMC studies, using plane-wave nodes, have been previously
performed~\cite{TC,rapisarda}.  According to Ref.~\cite{TC},
crystallization occurs directly from the PL, although, at freezing, the
energies of all three phases are very close to each other.  Subsequent
PW--based simulations\cite{kwon,rapisarda} revised slightly upwards
the energy of the PL, but this was enough to alter the previous
result, and to predict a small stability window for the 
FL\cite{rapisarda}.  Our backflow calculations lower back the energy of the
PL relative to the FL, with the effect of shrinking, but apparently
not eliminating, the density range where the FL phase is stable: the
ideal 2DEG undergoes a polarization transition at $r_s \sim 
26$, and the polarized liquid crystallizes at $r_s\sim 35$. 

{\it Acknowledgements --} 
We acknowledge partial financial support from MURST (the Italian 
Ministry for University, Research and Technology) through COFIN99.


\begin{thebibliography}{100}
\bibitem{nota}
We performed simulations for the solid phase at $r_s=30$ and 40. Our
results for $N=56$ particles agree with the ones of Ref.~\cite{rapisarda},
while showing small discrepancies with Ref.~\cite{TC}.
\end{thebibliography}
\end{document}